\documentclass[prl, floatfix,twocolumn, superscriptaddress]{revtex4}

\usepackage[utf8]{inputenc} 
\usepackage[T1]{fontenc}    
\usepackage{url}            
\usepackage{booktabs}       
\usepackage{amsfonts}       
\usepackage{amsmath}
\usepackage{nicefrac}       
\usepackage{lipsum}
\usepackage{graphicx}
\usepackage{placeins}
\usepackage{natbib}

\begin{document}
\title{Solitary routes to chimera states}

\author{Leonhard Sch\"ulen}
\affiliation{Institut f\"ur Theoretische Physik, Technische Universit\"at Berlin, Hardenbergstraße 36, 10623 Berlin}

\author{Alexander Gerdes}
\affiliation{Weierstrass Institute for Applied Analysis and Stochastics, Mohrenstrasse 39, 10117 Berlin, Germany}
 \author{Matthias Wolfrum}
\affiliation{Weierstrass Institute for Applied Analysis and Stochastics, Mohrenstrasse 39, 10117 Berlin, Germany}
\author{Anna Zakharova}
\affiliation{Institut f\"ur Theoretische Physik, Technische Universit\"at Berlin, Hardenbergstraße 36, 10623 Berlin}

\date{\today}
\begin{abstract}
We show how solitary states in a system of globally coupled FitzHugh-Nagumo oscillators can lead to the emergence of chimera states. By a numerical bifurcation analysis of a suitable reduced system in the thermodynamic limit we demonstrate how solitary states, after emerging from the synchronous state, become chaotic in a period-doubling cascade. Subsequently, states with a single chaotic oscillator give rise to states with an increasing number of incoherent chaotic oscillators. In large systems, these chimera states show extensive chaos. We demonstrate the coexistence of many of such chaotic attractors with different Lyapunov dimensions, due to different numbers of incoherent oscillators.
\end{abstract}
\maketitle
Solitary states in coupled oscillator systems  -- a counterpart to classical solitons in spatially extended systems -- are an interesting nonlinear pattern and have recently received much attention of researchers \cite{MAI2014,JAR18,SCH19,MAJ19,MIK19,BER20,KRU2020,SCH21,RYB21}.  They play an important role as a cornerstone to more complex self-organized states \cite{WOL15,RYB19,HEL20,FRA22}, e.g. in power grid models or neuronal systems. Here, we will use them to explain the origin of another intriguing nonlinear phenomenon in  coupled oscillator systems, namely the emergence of coherence/incoherence patterns, called {\em chimera states} \cite{ZAK20}. They are characterized as dynamical states, where in a  self-organized process a population of homogeneous oscillators splits into coherent and incoherent parts. Since their discovery \cite{KUR2002} it was a major open question how their emergence can be explained by a step-wise  supercritical scenario \cite{ABR2004}. Only recently, two results in this direction have been obtained. Haugland et al.  \cite{HAU2021} showed how they arise in a system with global non-linear coupling in a  cascade of cluster-splittings, after in \cite{SCH2015} clustering has been identified as a prerequisite for chimera states. 
In  \cite{FRA2021} Franovi\'{c} et al. showed a completely different scenario where in an array of excitable phase oscillators with attractive and repulsive coupling, coherence/incoherence patterns arise from a coherent Turing pattern by a homoclinic bifurcation with subsequent transition to extensive chaos. 

\begin{figure}[htbp]
    \centering
    \includegraphics[width=0.9\columnwidth]{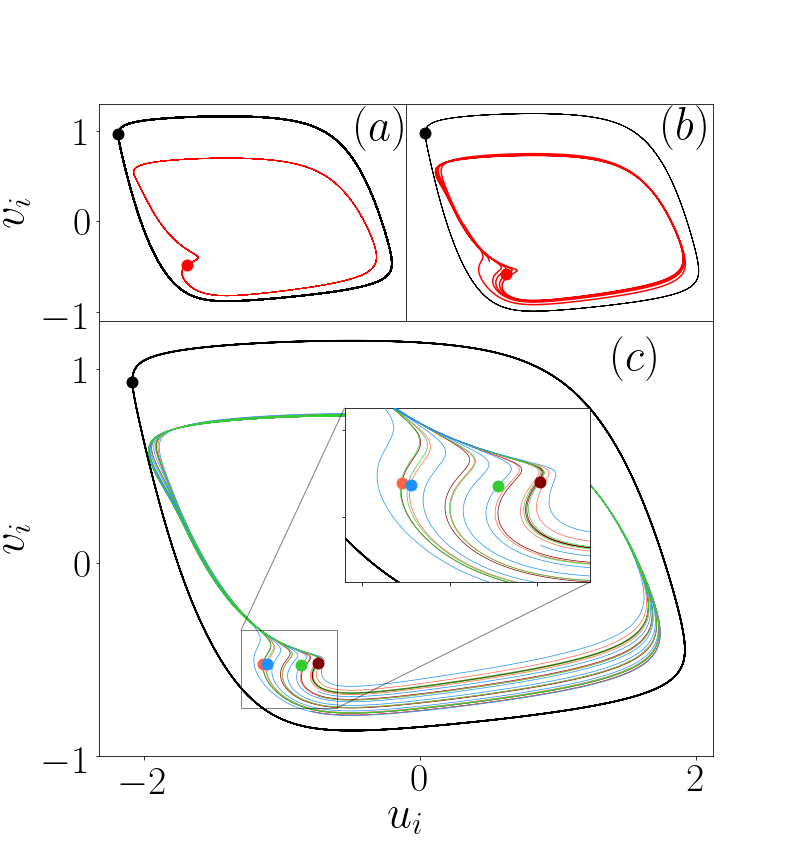}
    \caption{Phase portraits of different types of solutions for globally coupled FHN oscillators (\ref{Equ:FHN}). Trajectories of coherent (black) and incoherent (in color) oscillators. (a)  periodic solitary state at $\sigma_u=0.14$; (b) chaotic solitary state at $\sigma_u=0.12$; (c) chimera state with 4 incoherent oscillators at $\sigma_u=0.11$. Other parameters: $N=100$, $a=0.5$, $\varepsilon=0.1$, $\sigma_v=0.15$. All results were obtained from random  initial conditions, uniformly distributed in the intervals $u_i \in [-2.2,2.2]$ and $v_i \in [-1.2,1.2]$ for an integration time of $T=5000$ using the \emph{LSODA} method of the python package $scipy.integrate$ with an initial time-step of $dt=0.01$.}
    \label{fig:fig1}
\end{figure}

In this letter, we disclose another route to the emergence of chimera states.  We use a system of globally coupled FitzHugh-Nagumo (FHN) oscillators to demonstrate how solitary states can become an entry point to such patterns of localized extensive chaos. This transition occurs as follows (Fig.~\ref{fig:fig1}). For a fixed value of coupling strength parameter, the system demonstrates a periodic solitary state, where a single {\em solitary} oscillator performs an independent periodic motion (red (gray) trajectory in Fig.~\ref{fig:fig1}a), while all other oscillators form a stable synchronized cluster moving along the limit cycle of the FHN system in the oscillatory regime (black trajectory in Fig.~\ref{fig:fig1}a). Upon a variation of the coupling strength the temporal dynamics of the solitary oscillator becomes chaotic (Fig.~\ref{fig:fig1}b). For even lower coupling strenghts we  obtain solutions with several incoherent oscillators, each displaying an independent chaotic motion (Fig.~\ref{fig:fig1}c). The results in Fig.~\ref{fig:fig1} were obtained from random uniform initial conditions. Note that all states coexist with the stable fully synchronized solution and may coexist with stable solutions with other cluster types.
However, the solitary states in panels (a) and (b) are the most probable ones, when random initial conditions are chosen.In the parameter regime of the chimera state, shown in panel (c), we observe the coexistence of several similar states with different numbers of incoherent oscillators, which we will discuss below in more detail. 

Here, we provide a detailed study of this transition process. First, we use a thermodynamic limit description for a bifurcation analysis of the solitary states and demonstrate the transition to chaos in a classical period doubling cascade. Identifying in this way the parameter conditions and suitable initial conditions, we show how the chaotic solitary state gives rise to multiple  coexisting chimera states characterized by different numbers of incoherent oscillators. Based on a Lyapunov analysis, we show that they represent coexisting attractors with extensive chaos of different Lyapunov dimension.

Our model is a globally coupled system of  $N$ identical FHN oscillators: 
\begin{align}
    \varepsilon \frac{du_i}{dt} &= u_i - \frac{u_i^3}{3} - v_i + \sigma_u (\tilde u - u_i) + \sigma_v (\tilde v -v_i), \nonumber \\
    \frac{dv_i}{dt} &= u_i +a,  \quad \tilde u=\frac{1}{N} \sum_{j=1}^{N} u_j, \quad  \tilde v=\frac{1}{N} \sum_{j=1}^{N} v_j, \label{Equ:FHN}
\end{align}
$i=1,\dots ,N$ where $u_i$ and $v_i$ are the activator and inhibitor variables of the $i$-th oscillator, respectively, $\varepsilon$ determines the time scale separation between the fast ($u$) and the slow ($v$) variable. The strength of the coupling to the mean fields $\tilde u$ and $\tilde v$ is given by $\sigma_u$ and $\sigma_v$, respectively. Throughout the paper we fix  the threshold parameter $a=0.5$  in the oscillatory regime ($|a|<1$) far away from the Hopf bifurcation and use a moderate time scale separation $\varepsilon=0.1$.

\paragraph{Bifurcations of solitary states in the thermodynamic limit:}
Cluster states are self-organized patterns arising naturally in systems of identical units with global symmetric coupling. Each cluster type is characterized by a partition of the set of oscillators into subsets (clusters) with $u_i=u_j$ and $v_i=v_j$ whenever two oscillators $i$ and $j$ belong to the same cluster. By the symmetry of the system, this induces a corresponding dynamically invariant subspace, allowing a low-dimensional description of these states \cite{GOL2003}. The situation with only one cluster, i.e. all oscillators behaving identically, corresponds to (global) synchrony. The dynamics of a cluster state can be described by a reduced system within the invariant subspace with one pair of variables $u, v$ for each cluster and the cluster sizes represented by corresponding weights in the mean fields $\tilde u,\tilde v$.  Note that symmetry breaking bifurcations, which are transversal to the invariant subspace, are not covered by the reduced system. 

\begin{figure}
    \centering
    \includegraphics[width=0.95\columnwidth]{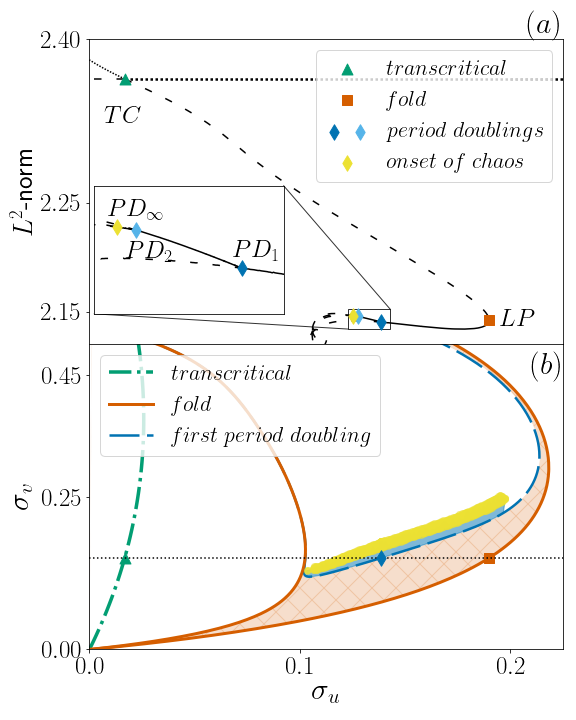}
    \caption{Bifurcations in the thermodynamic limit (\ref{Equ:thermo}). (a) Branches of synchronous and solitary states for varying $\sigma_u$ and fixed $\sigma_v=0.15$. 
    Synchronous branch (dotted horizontal line) with transcritical instability ($TC$ green triangle).
    Bifurcating solitary branch (stable/unstable parts are solid/dashed) with fold bifurcation (LP / red square), period doublings ($PD_1$ / dark and $PD_2$ / light blue (shaded) diamonds, see inset panel), and onset of chaos ($PD_\infty$ yellow (light shaded) diamond).
    (b) Parameter plane ($\sigma_u$,$\sigma_v$) with curves of fold (red solid), period-doubling (blue dashed), and transcritical bifurcations (green dash-dotted). Regions of stable solitary states (red hatched shading), period doubled solitary states (blue (light) shading) and chaotic solitary states (yellow (dark) shading). The densly dotted horizontal line indicates the $\sigma_v$ value of panel (a). 
    Other parameters are: $\varepsilon=0.1$, $a=0.5$.}
    \label{fig:fig2}
\end{figure}

\begin{figure}
    \centering
    \includegraphics[width=0.95\columnwidth]{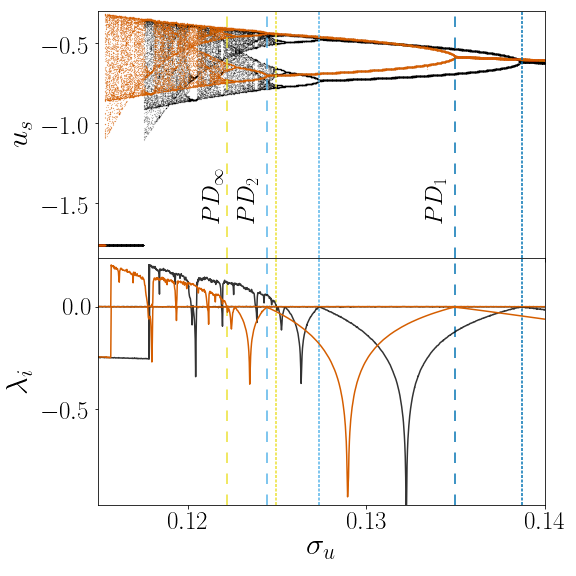}
    \caption{Period doubling cascade leading to a chaotic solitary state. Top: Sampled values of $u_s$ at Poincarè section $u_b=0$ for the thermodynamic limit $N=\infty$ (black) and for the finite size system with $N=100$ (red (gray)). Bottom: first and second Lyapunov exponent for $N=\infty$ (black) and $N=100$ (red (gray)). The vertical lines indicate the first two period doublings and the onset of chaos for the thermondynamic limit (dotted) and the finite size system (dashed, with names in top panel). Colors (shading) as in Fig. \ref{fig:fig2}. Attractor crisis at $\sigma_u \approx 1.1504$ ($N=\infty$) and $\sigma_u \approx 0.1153$ ($N=100$).}
    \label{fig:fig3}
\end{figure}

A specific type of cluster states are {\em solitary states}, where $N-1$ oscillators constitute one big cluster ("bulk"), while the remaining  single oscillator forms a (trivial) second cluster. For the thermodynamic limit of large system size $N \rightarrow \infty$, the mean fields $\tilde u,~\tilde v$ are equal to the  bulk variables $u_b, ~v_b$ and  we obtain 
\begin{align}
    \varepsilon \frac{du_b}{dt} &= u_b - \frac{u_b^3}{3} - v_b,  \qquad \frac{dv_b}{dt} = u_b +a, \nonumber \\
    \varepsilon \frac{du_s}{dt} &= u_s - \frac{u_s^3}{3} - v_s + \sigma_u (u_b - u_s) + \sigma_v (v_b-v_s), \nonumber \\
    \frac{dv_s}{dt} &= u_s +a,
    \label{Equ:thermo}
\end{align}
where the coupling term in the equations for the bulk variables vanishes. Therefore, the solitary oscillator can be interpreted as a probe particle driven by a mean-field to which its variables $u_s, v_s$ do not contribute. We will use this system to study the emergence of stable solitary states and their transition from a periodic to a chaotic regime. To this end we employ numerical bifurcation analysis based on path-following methods using the software $\mathbf{auto-07p}$ \cite{DOE07}. The bifurcation diagram in Fig.~\ref{fig:fig2}(a) shows a branch of  synchronous periodic states (dotted horizontal line) and a bifurcating branch of periodic solitary states (dashed and solid black curve) for varying $\sigma_u$ at fixed $\sigma_v=0.15$. The synchronous state does not depend on the coupling  $\sigma_u$, but its stability changes -- a well known phenomenon \cite{NAK1994} sometimes called Benjamin-Feir instability \cite{KEM21}. This instability manifests itself in the reduced system (Eqs. \ref{Equ:thermo}) as a transcritical bifurcation (TC / green triangle in Fig.~\ref{fig:fig2}a). The dashed horizontal branch left of this bifurcation corresponds to the now unstable synchronized solution, whereas the dotted upper left branch is stable only in the reduced system, but has no corresponding stable solution in the full system. The bifurcating branch of solitary states (black dashed curve) turns around in a fold bifurcation (SN / red square), where it gains stability and gives rise to stable periodic solitary states (black solid line). This stable branch undergoes a supercritical period doubling bifurcation (PD1 / dark blue (gray) diamond), where a stable branch of period doubled  solitary states (solid line in the inset) emerges. By a subsequent period-doubling cascade chaotic solutions similar to those shown in Fig. \ref{fig:fig1}(b) arise. 
Having obtained the bifurcation points for a fixed value of $\sigma_v$, we show in Fig.~\ref{fig:fig2}(b) the corresponding  bifurcation curves in the parameter plane ($\sigma_u$,$\sigma_v$). The green dash-dotted curve corresponds to the transcritical instability of the synchronous state.
The region of stable periodic solitary states (shaded, hatched region)
is bounded by the fold  bifurcation curve (red solid) and the period-doubling curve (blue dashed). Stable solitary states with higher periodicity are found in the rather small region (blue / light shaded), before chaos takes over (yellow / darker shaded). This region was obtained by scanning for a positive leading Lyapunov exponent. 
\paragraph{Chaotic solitary states:}
Next, we study the period doubling cascade of solitary states for a fixed value of $\sigma_v=0.15$ and decreasing coupling strength  $\sigma_u$ in the thermodynamic limit 
and for a finite size system with $N=100$. 
To this end, we analyze the sampled solution values $u_{s}$ at a suitable Poincaré section (Fig.~\ref{fig:fig3} top panel) and the two largest Lyapunov exponents (bottom panel). For the finite size system we observe a slight shift to lower values of $\sigma_u$. In both cases at a critical value of $\sigma_u$ the chaotic attractor collapses in an attractor crisis and the system falls back onto the  synchronous state. In order to validate the results of Fig. \ref{fig:fig2}, we indicate the values of $\sigma_u$ of the first and second period doubling and the onset of chaos both for the thermodynamic limit (densly dotted vertical lines) and the finite-size system (dashed vertical lines).

\begin{figure}
    \centering
    \includegraphics[width=0.99\columnwidth]{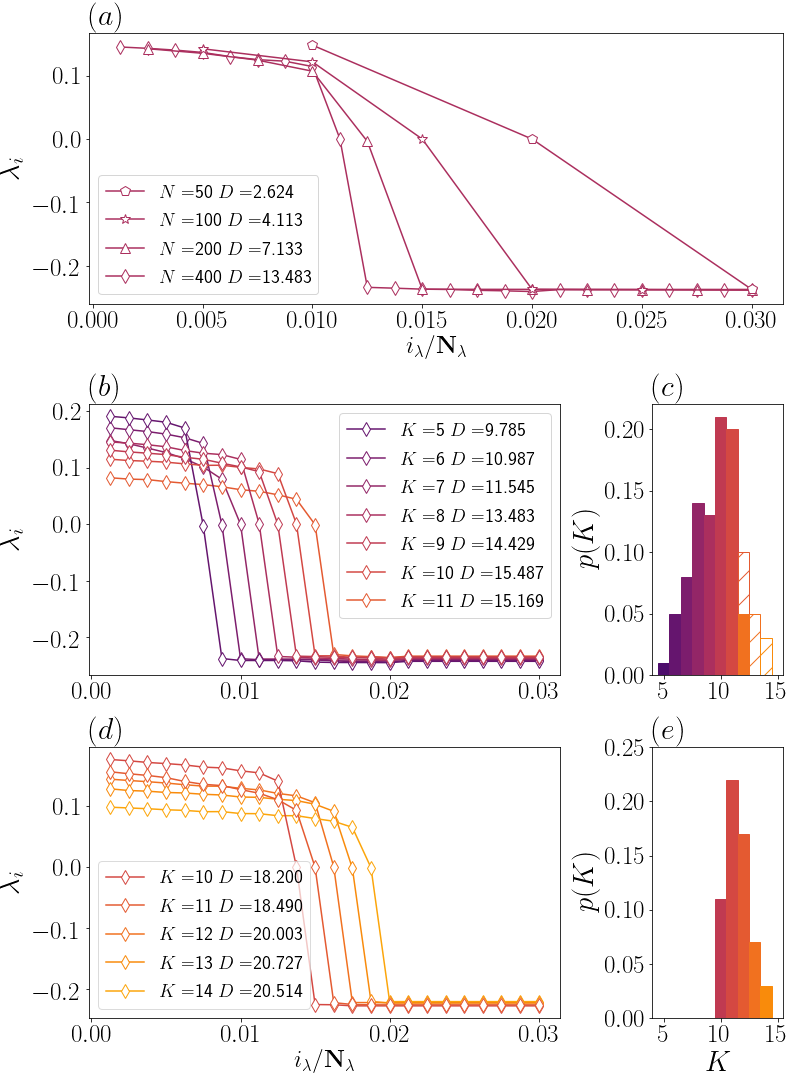}
    \caption{(a) Lyapunov spectra of chimera states demonstrating extensive chaos for various $N$ (see inset) and $K$ with fixed $K/N=0.02$. (b) Lyapunov spectra and attractor dimensions $D$ (see inset) of coexisting chimera states for $N=400$ and various $K$, $\sigma_u=0.115$. (c) Probability distribution $p(K)$ for $100$ random initial conditions, $\sigma_u=0.115$. Colored bins: chaotic chimera states, hatched: other states. (d), (e) show the same as (b),(c) for $\sigma_u=0.112$. Other paramters: $K_I=15$, $\varepsilon=0.1$, $a=0.5$.}
    \label{fig:fig4}
\end{figure}

\paragraph{Chimera states:}
In addition to the chaotic solitary states that we have established so far, where a single oscillator behaves chaotically and incoherent to the bulk, we demonstrate now solutions with more than one incoherent oscillator. We call a solution {\em chimera state} if all the oscillators apart from a large bulk cluster behave incoherent, i.e. all clusters except the bulk have size one. In this sense, a solitary state is a chimera state with only one incoherent oscillator. It turns out that chimera states appear in company with the chaotic solitary state and both the parameter values and initial conditions found in our bifurcation analysis of the thermodynamic limit are a good starting point to find them. 
Our strategy here is to pick a point $(\bar u_b, \bar v_b , \bar u_s, \bar v_s) = (-1.746619,-0.029879,-0.999828,-0.774970)$ 
on the chaotic solitary trajectory of the thermodynamic limit system, which we found for  $\sigma_u=0.118$,  $\sigma_v=0.15$ and generate an initial condition for a finite size system by initializing a large number $N-K$ of bulk oscillators at $(\bar u_b, \bar v_b )$. For the remaining $K$ potentially incoherent  oscillators we use independent small random perturbations  of $(\bar u_s + \delta, \bar v_s + \delta )$, equally distributed in a range of $\delta \in [-0.01,0.01]$. In order to compensate the shift in  $\sigma_u$ for finite size systems that we have noticed before (cf.  Fig.~\ref{fig:fig3}) we use in the simulations a smaller  value  $\sigma_u=0.115$. In this way we indeed obtain four different chimera trajectories in systems with size $N\in\{50,100,200,400\}$ and $K\in\{1,2,4,8\}$ incoherent oscillators, respectively. We observe that the solitary state with $N=50$ and $K=1$ upon doubling the system size induces  chimera states where the number $K$ of incoherent oscillators is doubled as well. In Fig.~\ref{fig:fig4}(a) we show the leading part of the Lyapunov spectra for these states. We observe that the number of positive exponents coincides with the number $K$ of incoherent oscillators. This extensive behavior for large $N$ is also reflected by the Lyapunov dimension, which we estimated by the 
Kaplan-Yorke formula. 
The fact that the dimensions, given in the figure legend, are always bigger than $K$ can be explained by the fact that each incoherent oscillator, performing an independent chaotic motion in the plane can make a contribution of sightly more than one to the total attractor dimension. This is different to the case of phase oscillators studied in \cite{WOL11}, where the Lyapunov dimension
almost exactly coincides with the number of incoherent oscillators.

\paragraph{Coexisting chimera states with different Lyapunov dimensions:}
In order to find coexisting chimera states with different numbers $K$ of incoherent oscillators, we repeated the numerical calculations for $N=400$ with a slightly different paradigm for the choice of the initial conditions. To allow also for a larger number of incoherent oscillators, we initialize only $N-K_I$ oscillators at the bulk values $(\bar u_b, \bar v_b ) = (-1.746619,-0.029879)$. For the remaining $K_I$ initially incoherent oscillators, we pick again random perturbations  of $(\bar u_s + \delta, \bar v_s + \delta )$, but now with the perturbations chosen equally distributed in a larger interval $\delta \in [-0.12,0.12]$. During a transient, which we took in our simulations as $T_t = 5000$, some of these initially incoherent oscillators will be absorbed by the bulk cluster. In some cases, they may also form small clusters, such that the final state is not a chimera state according to our definition above.  In most cases, however,  we obtain a chimera state, now with different numbers $K\in\{5,\dots , 11\}$ of incoherent oscillators. Note that our choice of $\sigma_u=0.115$  is already beyond the region of existence of the chaotic solitary state for the thermodynamic limit shown in Fig.~\ref{fig:fig3}, such that it is no surprise that we do not find a state with $K=1$ here. 

In Fig.~\ref{fig:fig4}(b) we show the leading part of the Lyapunov spectra for these coexisting states. Again, the number of positive exponents coincides with the number $K$ of incoherent oscillators. Whenever two different random initial conditions lead to the same $K$, we observe that the spectra and the corresponding Lyapunov dimensions coincide up to numerical accuracy. A histogram with the relative number of counts $p(K)$ is given in panel (c). Only in a small part  (the hatched part of the histogram)  of the resulting states are not chimera states. In these cases, some of the $K$ oscillators not belonging to the bulk form smaller clusters and the dynamics may be not chaotic, but periodic with a high period.
	
Fig.~\ref{fig:fig4}(d) and (e) show the same information for a slightly smaller parameter value $\sigma_u=0.112$. We see a qualitatively similar scenario where the number of incoherent oscillators  ranges in $K\in\{10,\dots ,14\}$ and we obtained chimera states for all initial conditions. We conclude that changing $\sigma_u$ towards smaller values shifts both the upper and the lower bound of possible numbers $K$ to larger values. Within this range we observe a unimodal  Gaussian-like distribution $p(K)$, compare \cite{WIL2006}, where a similar effect has been shown for coexisting twisted waves in a system of coupled phase oscillators. Interestingly, we see that the Lyapunov dimension may even decrease towards larger $K$, indicating that each single incoherent oscillator behaves "less chaotic" close to the upper bound of possible $K$ and hence the total dimension may decrease for an increasing number of incoherent oscillators. 

All these coexisting chimera states can, in principle, be found from initial conditions  chosen completely randomly, as presented in Fig.~\ref{fig:fig1}. However, exploring fully this rich scenario of coexisting states of different types, some of them with very small basins of attraction and hard to find from random initial conditions, goes beyond the scope of this paper, where we decided to focus our attention on the emergence of chimera states and their coexistence.

\paragraph{Conclusion and outlook:} 
While it is well known that self-organized wave patterns typically coexist within an interval of possible different wave numbers (Busse ballon \cite{SLB65}, Eckhaus stability region \cite{Eckhaus1965,BarT1990}), and also regular cluster solutions in globally coupled oscillator systems coexist for different cluster sizes \cite{Okuda1993,HMM1993}, we show here the coexistence of coherence-incoherence patterns with different numbers of incoherent oscillators, which are in fact coexisting chaotic attractors with different Lyapunov dimensions. The incoherent oscillators in these coexisting attractors show extensive chaos of different dimensions. The total share of incoherent oscillators in a chimera state is a macroscopic quantity. Hence, within the range of such shares, where stable chimera states exist, we find, 
for large systems, an increasing number of coexisting attractors with their numbers of incoherent oscillators increasing as well. We showed that, varying the coupling parameter, this extensive scenario is linked to the thermodynamic limit of the solitary regime, where the range of admissible numbers of incoherent oscillators shrinks down to one single oscillator in an infinitely large system. For this case, the emergence of the chaotic motion of the single incoherent oscillator could be shown in a period doubling cascade. 

\paragraph{Acknowledgements}
We thank Everton S. Medeiros for fruitful discussions.
This work was supported by the Deutsche Forschungsgemeinschaft (DFG) within the framework of the SFB 910 - Project number 163436311.

\bibliographystyle{unsrt}  
\bibliography{references}  

\begin{thebibliography}{10}

\bibitem{MAI2014}
Y.~Maistrenko, B.~Penkovsky, and Rosenblum M.
\newblock Solitary state at the edge of synchrony in ensembles with attractive
  and repulsive interactions.
\newblock {\em Phys. Rev. E}, 89:060901, 2014.

\bibitem{JAR18}
P.~Jaros, S.~Brezetsky, R.~Levchenko, D.~Dudkowski, T.~Kapitaniak, and
  Y.~Maistrenko.
\newblock Solitary states for coupled oscillators with inertia.
\newblock {\em Chaos}, 28(1):011103, 2018.

\bibitem{SCH19}
L.~Schülen, S.~Ghosh, A.D. Kachhvah, A.~Zakharova, and S.~Jalan.
\newblock Delay engineered solitary states in complex networks.
\newblock {\em Chaos, Solitons \& Fractals}, 128:290 -- 296, 2019.

\bibitem{MAJ19}
S.~Majhi, T.~Kapitaniak, and D.~Ghosh.
\newblock Solitary states in multiplex networks owing to competing
  interactions.
\newblock {\em Chaos}, 29(1):013108, 2019.

\bibitem{MIK19}
M.~Mikhaylenko, L.~Ramlow, S.~Jalan, and A.~Zakharova.
\newblock Weak multiplexing in neural networks: Switching between chimera and
  solitary states.
\newblock {\em Chaos}, 29(2):023122, 2019.

\bibitem{BER20}
R.~Berner, A.~Polanska, E.~Sch{\"o}ll, and S.~Yanchuk.
\newblock Solitary states in adaptive nonlocal oscillator networks.
\newblock {\em The European Physical Journal Special Topics},
  229(12):2183--2203, 2020.

\bibitem{KRU2020}
N.~Kruk, Y.~Maistrenko, and H.~Koeppl.
\newblock Solitary states in the mean-field limit.
\newblock {\em Chaos}, 30(11):111104, 2020.

\bibitem{SCH21}
L.~Schülen, D.A. Janzen, E.S. Medeiros, and A.~Zakharova.
\newblock Solitary states in multiplex neural networks: Onset and
  vulnerability.
\newblock {\em Chaos, Solitons \& Fractals}, 145:110670, 2021.

\bibitem{RYB21}
E.V. Rybalova, A.~Zakharova, and G.I. Strelkova.
\newblock Interplay between solitary states and chimeras in multiplex neural
  networks.
\newblock {\em Chaos, Solitons \& Fractals}, 148:111011, 2021.

\bibitem{WOL15}
M.~Wolfrum.
\newblock The turing bifurcation in network systems: Collective patterns and
  single differentiated nodes.
\newblock {\em Physica D: Nonlinear Phenomena}, 241(16):1351 -- 1357, 2012.

\bibitem{RYB19}
E.~Rybalova, V.~S. Anishchenko, G.~I. Strelkova, and A.~Zakharova.
\newblock Solitary states and solitary state chimera in neural networks.
\newblock {\em Chaos}, 29(7):071106, 2019.

\bibitem{HEL20}
F.~Hellmann, P.~Schultz, P.~Jaros, R.~Levchenko, T.~Kapitaniak, J.~Kurths, and
  Y.~Maistrenko.
\newblock Network-induced multistability through lossy coupling and exotic
  solitary states.
\newblock {\em Nature Communications}, 11(592), 2020.

\bibitem{FRA22}
I.~Franovi{\'c}, S.~Eydam, N.~Semenova, and A.~Zakharova.
\newblock Unbalanced clustering and solitary states in coupled excitable
  systems.
\newblock {\em Chaos}, 32(1):011104, 2022.

\bibitem{ZAK20}
A.~Zakharova.
\newblock {\em Chimera Patterns in Networks: Interplay between Dynamics,
  Structure, Noise and Delay}.
\newblock Springer, 2020.

\bibitem{KUR2002}
Y.~Kuramoto and D.~Battogtokh.
\newblock Coexistence of coherence and incoherence in nonlocally coupled phase
  oscillators.
\newblock {\em Nonlinear Phenomena in Complex Systems}, 4(5):380--385, 2002.

\bibitem{ABR2004}
D.~M. Abrams and S.~H. Strogatz.
\newblock Chimera states for coupled oscillators.
\newblock {\em Phys. Rev. Lett.}, 93:174102, Oct 2004.

\bibitem{HAU2021}
S.~W. Haugland, A.~Tosolini, and K.~Krischer.
\newblock Between synchrony and turbulence: intricate hierarchies of
  coexistence patterns.
\newblock {\em Nature Communications}, 12(1):5634, 2021.

\bibitem{SCH2015}
L.~Schmidt and K.~Krischer.
\newblock Clustering as a prerequisite for chimera states in globally coupled
  systems.
\newblock {\em Phys. Rev. Lett.}, 114:034101, Jan 2015.

\bibitem{FRA2021}
I.~Franovi\ifmmode~\acute{c}\else \'{c}\fi{}, O.~E. Omel'chenko, and
  M.~Wolfrum.
\newblock Bumps, chimera states, and turing patterns in systems of coupled
  active rotators.
\newblock {\em Phys. Rev. E}, 104:L052201, Nov 2021.

\bibitem{GOL2003}
M.~Golubitsky and I.~Stewart.
\newblock {\em The symmetry perspective: from equilibrium to chaos in phase
  space and physical space}, volume 200.
\newblock Springer Science \& Business Media, 2003.

\bibitem{DOE07}
E.~J. Doedel, A.~R. Champneys, F.~Dercole, T.~F. Fairgrieve, Y.~A. Kuznetsov,
  B.~Oldeman, R.C. Paffenroth, B.~Sandstede, X.J. Wang, and C.H. Zhang.
\newblock Auto-07p: Continuation and bifurcation software for ordinary
  differential equations.
\newblock 2007.

\bibitem{NAK1994}
N.~Nakagawa and Y.~Kuramoto.
\newblock From collective oscillations to collective chaos in a globally
  coupled oscillator system.
\newblock {\em Physica D: Nonlinear Phenomena}, 75(1-3):74--80, 1994.

\bibitem{KEM21}
F.~P. Kemeth, B.~Fiedler, S.~W. Haugland, and K.~Krischer.
\newblock 2-cluster fixed-point analysis of mean-coupled stuart--landau
  oscillators in the center manifold.
\newblock {\em Journal of Physics: Complexity}, 2(2):025005, 2021.

\bibitem{WOL11}
M.~Wolfrum and E.~Omel’chenko.
\newblock Chimera states are chaotic transients.
\newblock {\em Physical Review E}, 84(1):015201, 2011.

\bibitem{WIL2006}
D.~A. Wiley, S.~H. Strogatz, and M.~Girvan.
\newblock The size of the sync basin.
\newblock {\em Chaos}, 16(1):015103, 2006.

\bibitem{SLB65}
A.~Schlüter, D.~Lortz, and F.~Busse.
\newblock On the stability of steady finite amplitude convection.
\newblock {\em Journal of Fluid Mechanics}, 23(1):129–144, 1965.

\bibitem{Eckhaus1965}
W.~Eckhaus.
\newblock {\em Studies in Non-Linear Stability Theory}.
\newblock Springer Tracts in Natural Philosophy. Springer, Berlin, Heidelberg,
  1965.

\bibitem{BarT1990}
L.~Tuckerman and D.~Barkley.
\newblock Bifurcation analysis of the eckhaus instability.
\newblock {\em Physica D: Nonlinear Phenomena}, 46:57--86, 10 1990.

\bibitem{Okuda1993}
K.~Okuda.
\newblock Variety and generality of clustering in globally coupled oscillators.
\newblock {\em Physica D: Nonlinear Phenomena}, 63(3):424–436, 1993.

\bibitem{HMM1993}
D.~Hansel, G.~Mato, and C.~Meunier.
\newblock Clustering and slow switching in globally coupled phase oscillators.
\newblock {\em Phys. Rev. E}, 48:3470, 1993.

\end{thebibliography}

\end{document}